\documentclass[11pt,letterpaper]{article}
\usepackage{mathtools}
\usepackage{graphicx}
\usepackage{tikz}
\usepackage{tikz-feynman}
\usepackage{amscd}
\usepackage{amsmath}
\usepackage{xkcdcolors}
\usepackage{slashed}
\usepackage{gensymb}
\usepackage{bm}
\usepackage{xspace}
\usepackage{jheppub} 
\usepackage{graphicx}

\newcommand{\ZP}{\ensuremath{Z^\prime}\xspace}
\newcommand{\MZP}{\ensuremath{M_{Z^\prime}}\xspace}
\newcommand{\GZP}{\ensuremath{g_{Z^\prime}}\xspace}

\newcommand{\herwig}{H\protect\scalebox{0.8}{ERWIG}\xspace}
\newcommand{\rivet}{R\protect\scalebox{0.8}{IVET}\xspace}

\newcommand{\contur}{\textsc{Contur}\xspace}

\newcommand{\model}{Plan B}

\newcommand{\HLLHC}{2.4~TeV}

\usepackage{tikz}

\DeclareRobustCommand{\swatch}[1]{\tikz[baseline=-0.6ex]\node[fill=#1,shape=rectangle,draw=black,thick,minimum width=5mm,rounded corners=0.5pt](){};}
\newcommand{\MET}{\ensuremath{E_T^{\rm miss}}}

\definecolor{turquoise}{HTML}{40E0D0}
\definecolor{blue}{HTML}{0000FF}
\definecolor{salmon}{HTML}{FA8072}
\definecolor{darkgoldenrod}{HTML}{B8860B}
\definecolor{orange}{HTML}{FFA500}

\title{\centering 
  The Plan B Model: ${\bm Z^\prime}$ collider phenomenology and discovery prospects}

\author[a]{Ben Allanach,}
\affiliation[a]{DAMTP, University of Cambridge, Wilberforce Road, Cambridge, 
  CB3 0WA, United Kingdom}
\emailAdd{B.C.Allanach@damtp.cam.ac.uk}

\author[a,b,1]{Hannah Banks\note{Corresponding author.}}
\affiliation[b]{Center for Cosmology and Particle Physics, Department of Physics, New York University, New York, NY 10003, USA}
\emailAdd{hannah.banks@nyu.edu}

\author[c]{and Jon Butterworth}
\affiliation[c]{Department of Physics and Astronomy, University College London, Gower
  Street, London WC1E 6BT, United Kingdom}
\emailAdd{j.butterworth@ucl.ac.uk}

\abstract{The Plan B Model was proposed in
  Ref.~\cite{Allanach:2023uxz} and explains some gross features of the
  fermion mass spectrum.
  It also affects the predictions
  of various observables involving the $b \rightarrow s$ quark-flavour
  transition.
  The model predicts a new~TeV-scale \ZP which decays into various
  different final states. 
  We constrain the viable parameter space of the model by re-casting
  several LHC direct searches for the  \ZP in addition to using  \contur to 
  test the model against unfolded measurements of various differential 
  cross sections that are predicted to be non-zero in the Standard Model.
  We delineate the regions of parameter space that are excluded by current LHC data, 
  and estimate the projected reach of the high luminosity LHC. 
}

\begin{document} 
\maketitle
\flushbottom

\section{Introduction \label{sec:intro}}
Searches at the LHC for new, previously undiscovered particles often take the
form of bump-hunts on relatively smooth backgrounds. Indeed, the Higgs boson
was originally discovered in direct searches in this manner~\cite{ATLAS:2012yve,CMS:2012qbp}. As collider experiments push to ever higher centre-of-mass energies, quick gains can often be
obtained by performing a rough-and-ready search in the context of simplified models.
However, for a more specific model such as the \model\ Model proposed in Ref.~\cite{Allanach:2023uxz},
it is also important to cross-check against the growing library of measurements made at
comparable energies.
Here, we shall place bounds upon a new particle, the \ZP of the \model\ Model, by re-casting
various LHC bump-hunts, while also checking consistency against several LHC differential cross-section measurements
on which the model could in principle have an impact.

The \model\ Model augments the gauge symmetry of the Standard Model (SM) with an additional
anomaly-free $U(1)_X$ gauge group, under which the  fermions are assigned charges according
to thrice third-family baryon number minus electron number minus twice
muon number (cf.\ Table~\ref{tab:charges}), i.e.
\begin{equation}
  X = 3B_3 - L_1 -2L_2.
\end{equation}
The chiral fermionic field content of the model is identical to
that of the SM plus three right-handed neutrinos, which may generically account for
neutrino masses and mixing although the details of this within the \model\ Model have yet to be explored.
The $U(1)_X$ gauge symmetry is assumed to be spontaneously broken by a unit $X$-charged complex scalar SM-singlet
field $\theta$ (the `flavon') with a~TeV-scale vacuum expectation value $\langle \theta \rangle$, resulting in a
TeV-scale electrically neutral spin-1 vector boson dubbed the \ZP. Its
mass is given by
\begin{equation}
  M_{Z^\prime} = g_{Z^\prime} \langle \theta \rangle,
\end{equation}
where $g_{Z^\prime}$ is the $U(1)_X$ gauge coupling.
The model explains at a gross level why the CKM elements $|V_{ts}|$, $|V_{cb}|$, $|V_{ub}|$,
$|V_{td}|$ are all small: they are predicted to vanish in the limit of
unbroken $U(1)_X$ symmetry, but are expected to receive small corrections from
its spontaneous breaking.
The \model\ Model can
alter the predictions for $B$-meson observables involving the $b \rightarrow s
l^+ l^-$ transition (and its $CP$-conjugate) via interference between the SM
contribution and a new physics contribution mediated by the \ZP, which
couples to $l^+ l^-$ and $b \overline{s}+ h.c$. 
When fitted -- with two additional effective parameters -- to hundreds of measurements
including differential LEP di-lepton production cross sections and
measurements of $B$-decays and $B$-mixing, an improvement\footnote{The
anomalous magnetic moment of the muon is not included in the $\chi^2$ (but
receives a negligible \ZP contribution in the parameter space of
interest). Neutrino trident production $(\nu_\mu N \rightarrow \nu_\mu N
\mu^+\mu^-)$, for nucleus $N$, is also not included in the $\chi^2$, however the corresponding constraint, $M_{Z^\prime}/g_{Z^\prime} \geq 1.1$~TeV~\cite{Altmannshofer:2014pba}, is easily
satisfied throughout the parameter space of interest.}
of 34 units of $\chi^2$ was
observed~\cite{Allanach:2023uxz}. Much of the $\chi^2$ improvement comes from
observables involving the $b \rightarrow s \mu^+ \mu^-$ transition (or
$CP$-conjugate), notably differential measurements of $BR(B_s \rightarrow \phi
\mu^+ \mu^-)$~\cite{LHCb:2021zwz,CDF:2012qwd}, differential branching
ratio and angular distributions in $B\rightarrow K^\ast \mu^+ \mu^-$ decays~\cite{LHCb:2013ghj,LHCb:2015svh,ATLAS:2018gqc,CMS:2017rzx,CMS:2015bcy,Bobeth:2017vxj} and differential branching ratio measurements of
$B\rightarrow K \mu^+ \mu^-$~\cite{Parrott:2022zte}. The SM predictions of such observables contain
rather large theoretical uncertainties, which were estimated and taken into
account in the fit, although controversy about their reliability
remains~\cite{Ciuchini:2022wbq,Gubernari:2022hxn}. Many
other measurements, for example those involving lepton flavour universality
ratios and $B_s-\overline{B_s}$ mixing, roughly agree with their SM
predictions~\cite{LHCb:2022vje,LHCb:2022qnv} and were also taken into account in the fit.

\begin{table}
  \begin{center} 
    \begin{tabular}{|c|c|c|c|c|c|c|} \hline
        Family $i$ & $Q_i$ & $L_i$ & $u_i$ & $d_i$ & $e_i$ & $\nu_i$ \\ \hline
        $1$ & $0$ & $-1$ & $0$ & $0$ & $-1$ & $-1$ \\
        $2$ & $0$ & $-2$ & $0$ & $0$ & $-2$ & $-2$ \\
        $3$ & $1$ & $0$ & $1$ & $1$ & $0$ & $0$ \\        
        \hline
    \end{tabular}
    \caption{\label{tab:charges} $X$-charges of fermionic chiral fields
      under the $U(1)_X$ gauge group. $i \in \{1,2,3\}$ is a family
      index and gauge indices have been suppressed. $Q_i$ and $L_i$ are
      left-handed Weyl fermions, whereas the other fields listed are all
      right-handed Weyl fermions. The fields' representations under the SM
      $(SU(3),\ SU(2),\ U(1)_Y)$ gauge Lie algebra are:
      $Q_i \sim (3, 2, 1/6)$,
      $L_i \sim (1, 2, -1/2)$,
      $u_i \sim (3, 1, 2/3)$,
      $d_i \sim (3, 1, -1/3)$,
      $e_i \sim (1, 1, -1)$,
      $\nu_i \sim (1, 1, 0)$.        }
    \end{center}
  \end{table}

The \model\ Model's \ZP could show up as a resonance in various LHC
searches: $l^+ l^-$, di-jet, di-bottom and di-top searches (occasionally with
an additional bottom- or anti-bottom jet). 
The ATLAS and CMS experiments at the LHC have carried out such searches and
have yet to find convincing evidence of such a resonance. 
One may thus recast the direct resonance searches from the LHC in order to
bound the parameter space of the \model\ Model.
A line in $X$-charge space which includes the \model\ Model was examined in
Ref.~\cite{Allanach:2024jls} in order to recast bounds from
an ATLAS di-muon resonance search~\cite{ATLAS:2019erb}. 
It was concluded that $M_{Z^\prime}>1.2$~TeV from this search, when the other
parameters are adjusted to fit the $B$ meson decay and oscillation measurements
(and LEP2 measurements of differential di-lepton production
cross sections) as well as possible. We shall henceforth refer to the values
of these
other parameters as the best-fit indirect-measurement parameters (IMPs).
It was 
estimated that for the best-fit
IMPs, the high-luminosity run of
the LHC (HL-LHC) might have sensitivity up to $M_\ZP \leq $ \HLLHC\
assuming 3000 fb$^{-1}$ of integrated luminosity ~\cite{Allanach:2024jls}. 
Note that to
determine the fit IMPs of the \model\ Model Ref.~\cite{Allanach:2023uxz} used the language of the SM effective
field theory (SMEFT)~\cite{Grzadkowski:2010es}. Because the fit used
LEP2 di-lepton production and electroweak data,
we require $M_{Z^\prime} > 400$~GeV for the
SMEFT to be a valid approximation in determining such measurements. 

We extend the previous work in two ways: by studying the full 95$\%$ preferred
region rather than 
the best-fit
IMPs and also by considering other search channels and experiments. 
Instead of choosing the best-fit IMPs, we shall stay within the 95$\%$ region of
\emph{preferred} fit in the
\model\ Model (we refer to this region of parameter space as the preferred-fit IMPs).
As well as the pure di-lepton bump hunts considered before, 
we compute
constraints on the parameter space of the
\model\ Model from channels with additional $b$-jets
or $b$-vetoes
(the most important turn out to be di-electrons and di-muons),
in order to bound the parameter
space of interest
in the most robust manner possible. In particular, a recent di-muon plus
additional $b$-jets search by the CMS 
experiment, particularly designed to test models like the
\model\ Model~\cite{CMS:2023byi}, is employed in \S\ref{sec:cmsdimuonb}.
We shall also bound the preferred-fit IMPs 
by cross-checking against a broad spectrum of LHC measurements for
which the predicted SM cross section is substantial, but relatively well
known, with the \contur tool~\cite{Butterworth:2016sqg,Buckley:2021neu}
in \S\ref{sec:obs}.
We also provide rough sensitivity estimates for the HL-LHC\@.
These allow an identification (within the
currently allowed parameter space forecast to be accessible at the HL-LHC) of which
channels and experimental analyses beyond the more standard $\mu^+\mu^-$ searches,  have discovery potential for the \model\ Model's \ZP.  Indeed,  since the relative rates of different channels are predicted at each point in parameter space, measuring  multiple channels is important for model testing. 

The rest of the paper proceeds as follows: in \S\ref{sec:model}
the \model\ Model is introduced and the parameters that are important for the present paper are
introduced. In \S\ref{sec:recast}, we describe the methodology for our
calculation of \ZP search limits and sensitivity estimates and then
proceed to
present the resulting constraints upon the \model\ Model's parameter space.
We go on to check constraints coming from measurements at the LHC of differential 
cross sections in a variety of final-state observables
in \S\ref{sec:obs}, before providing 
a summary and a discussion in \S\ref{sec:conc}. 

\section{The \model\ Model \label{sec:model}}
\subsection{Definition and \ZP couplings}
Following Ref.~\cite{Allanach:2023uxz}, we write the fermionic fields in the gauge eigenbasis with a
primed notation
\begin{eqnarray}
{\bf u_L'}&=&\left( \begin{array}{c} u_L' \\ c_L' \\ t_L' \\ \end{array}
\right), \qquad
{\bf d_L'}=\left( \begin{array}{c} d_L' \\ s_L' \\ b_L' \\ \end{array}
\right), \qquad
{\bf e_L'}=\left( \begin{array}{c} e_L' \\ \mu_L' \\ \tau_L' \\ \end{array}
\right), \qquad
{\bm \nu_L'}=\left( \begin{array}{c} {\nu_e'}_L \\ {\nu_\mu'}_L
  \\ {\nu_\tau'}_L \\ \end{array} \right),
 \nonumber \\ 
{\bf u_R'}&=&\left( \begin{array}{c} u_R' \\ c_R' \\ t_R' \\ \end{array}
\right), \qquad
{\bf d_R'}=\left( \begin{array}{c} d_R' \\ s_R' \\ b_R' \\ \end{array}
\right),\qquad
{\bf e_R'}=\left( \begin{array}{c} e_R' \\ \mu_R' \\ \tau_R' \\ \end{array}
\right), \qquad
{\bm \nu_R'}=\left( \begin{array}{c} {\nu_e'}_R \\ {\nu_\mu'}_R
  \\ {\nu_\tau'}_R \\ \end{array} \right),
\end{eqnarray}
along with the SM fermionic electroweak doublets
\begin{equation}
{\bf Q'}_i=\left( \begin{array}{c} {\bf u_L'}_i \\ {\bf d_L'}_i \end{array}
\right),\qquad
{\bf L'}_i=\left( \begin{array}{c} {\bm \nu_L'}_i \\ {\bf e_L'}_i \end{array}
\right).  
\end{equation}
We remark that the model has a chiral fermion content equivalent to that of
the SM augmented by three right-handed SM-singlet fields, which could play the
role of right-handed neutrinos.
The \model\ Model is intended to be a valid description of the effective quantum field
theory at the~TeV scale, but remains incomplete. While one of the right-handed
neutrinos could be identified as a dark matter candidate~\cite{Dodelson:1993je} and leptogenesis~\cite{Fukugita:1986hr}
could potentially occur from decays of the heavier right-handed neutrinos, the
Yukawa sector appears to have additional structure in it such as the suppression of
the lighter fermion families' masses. Such structure could come from
higher scales (from further broken family symmetries, for instance). In our
bottom-up approach to model building, we consign such further higher-scale model building
to future endeavour. If direct evidence in support of the \model\ Model is
forthcoming from collider searches, it may then be timely to consider further
extensions to address this. 

The
{\em non}-primed {\em mass}\/ eigenstates\footnote{${\bf P}$ and ${\bf P}'$
are column 3-vectors in family space.} are identified as
\begin{equation}
  {{\bf P}} = V_P^\dag {\bf P'} \ \text{where}\ {P} \in \{{u_R},\
  {d_L},\ {u_L},\ {e_R},\ {\nu_R},\ {d_R},\ {\nu}_{L},\ {e_L}\}. \label{fermion_rotations}  
  \end{equation}
The model has strong assumptions about $V_P$:
$V_{e_R}=V_{d_R}=V_{u_R}=V_{e_L}=I$, the 3 by 3 identity matrix, 
\begin{equation}
  V_{d_L} = \left(\begin{array}{ccc} 1 & 0 & 0 \\
  0 & \cos \theta_{sb} & \sin \theta_{sb} \\
  0 & -\sin \theta_{sb} & \cos \theta_{sb} \\ \end{array}
  \right), \label{ansatz}
  \end{equation}
$V_{u_L}=V_{d_L} V^\dag$, $V_{\nu_L} = V_{e_L} U^\dag$, respectively, where
  $V$ is the CKM matrix and $U$ is the PMNS matrix (both taken to be the 
  central values in the standard parameterisation taken from
  Ref.~\cite{ParticleDataGroup:2022pth}).
  Note that $V_{e_L}=I$ means that there is no tree-level charged
  lepton-flavour violation and so $\mu \rightarrow e \gamma$ processes should
  impose no constraint. 

The couplings of the \ZP to the fermionic fields' mass eigenstates are
described by the Lagrangian density terms
\begin{eqnarray}
{\mathcal L} _I&=& - g_{Z^\prime} \left(\overline{\bf u_L} \slashed{Z}^\prime \Lambda_\xi^{u_L}{\bf
  u_L} +
\overline{\bf d_L} \slashed{Z}^\prime\Lambda_\xi^{d_L} {\bf d_L} +
  \overline{\bf u_R} \slashed{Z}^\prime\Lambda_\xi^{u_R} {\bf u_R} +
  \overline{\bf d_R} \slashed{Z}^\prime\Lambda_\xi^{d_R} {\bf d_R} +
  \right. \nonumber \\ && \left.
  \overline{\bf e_L} \slashed{Z}^\prime\Lambda_\Xi^{e_L} {\bf e_L} +
  \overline{\bf e_R} \slashed{Z}^\prime\Lambda_\Xi^{e_R} {\bf e_R} +
  \overline{\bm \nu_L} \slashed{Z}^\prime\Lambda_\Xi^{\nu_L} {\bm \nu_L} +
  \overline{\bm \nu_R} \slashed{Z}^\prime\Lambda_\Xi^{\nu_R} {\bm \nu_R}
  \right), \label{intsMass}  
\end{eqnarray}
where $\Lambda_{\alpha}^{P}:=V_{P}^\dag \alpha V_{P}$ for 
$\alpha \in \{ \xi, \Xi \}$ and
where
\begin{equation}
  \xi := \begin{pmatrix}
    0 & 0 & 0 \\
    0 & 0 & 0 \\
    0 & 0 & 1 \\
  \end{pmatrix}, \qquad
  \Xi := \begin{pmatrix}
    -1 & 0 & 0 \\
    0 & -2 & 0 \\
    0 & 0 & 0 \\
  \end{pmatrix}~,
\end{equation}
are fixed by the charges in Table~\ref{tab:charges}.
The right-handed neutrinos ${\bm \nu'_R}$ are assumed to be heavy compared to
the~TeV scale and play no further role in the LHC collider phenomenology. 

\subsection{Phenomenology}

In a global fit to 395 flavour, LEP2 and other measurements, the
\model\ Model's best-fit parameters come out to be~\cite{Allanach:2023uxz}:
\begin{equation}
  g_{Z^\prime}^{BF}\times \frac{\text{3~TeV}}{M_{Z^\prime}} = 0.206, \qquad
  \theta_{sb}^{BF} = -0.0406, \label{BFPars}
\end{equation}
  with an improvement in the
  total $\chi^2$ of 34 units.
These parameters can vary; they are
\begin{equation}
0.03\times \frac{M_{Z^\prime}}{\text{1~TeV}}  \leq g_{Z^\prime}\leq 0.15\times \frac{M_{Z^\prime}}{\text{1~TeV}}, \qquad
  -0.15 < \theta_{sb} < -0.02, \label{rangePars}
  \end{equation}
within 95$\%$ confidence level (CL) limits of the global fit.

\begin{figure}
    \begin{center}
  
      \includegraphics[width=0.98 \textwidth]{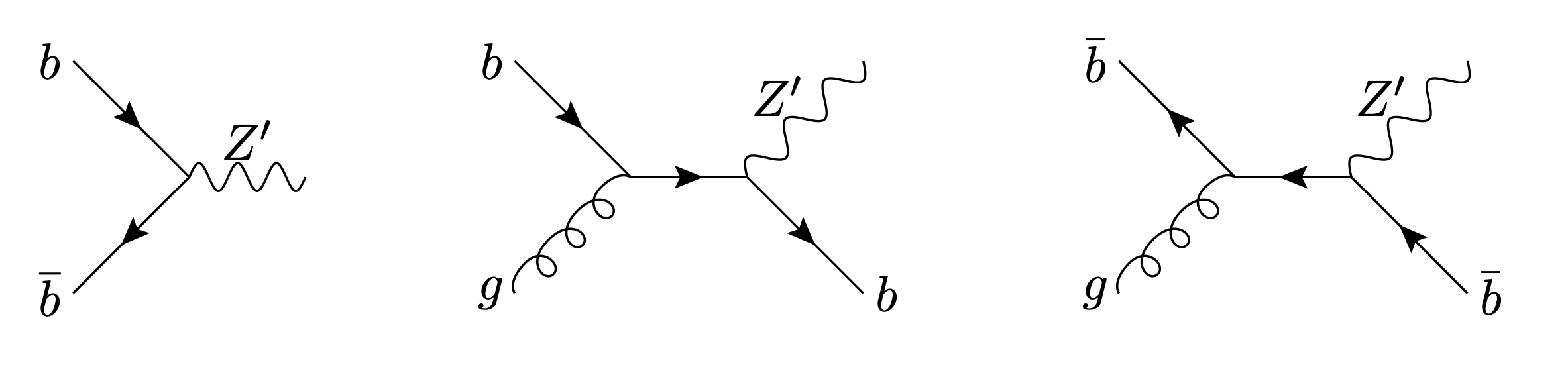}

      \caption{\label{fig:ffbar}
  LHC production of a \ZP boson. In the middle panel, we
show the associated production with a $b$ quark and in the right-hand panel,
the associated production with an anti-$\bar b$ quark.}
      \end{center}
      \end{figure}

LHC \ZP production cross section $\sigma$ in this model is dominated by the $b \bar b
\rightarrow Z^\prime$ process, which is doubly suppressed by the small bottom
and anti-bottom parton distribution functions. This contribution  is
\begin{equation}
  \sigma_{b b}\propto g_{Z^\prime}^2 (1 + \cos^4 \theta_{sb}) = 2 g_{Z^\prime}^2
  \left( 1 -
  \theta_{sb}^2 + {\mathcal{O}}(\theta_{sb}^4) \right), \label{approx}
\end{equation}
where the $1$ in the first parenthesis comes from the un-mixed right-handed bottom
quark contribution.
The next largest contribution $\sigma_{sb} \propto g_{Z^\prime}^2 \sin^2
2\theta_{sb}$ is from $s \bar b$ fusion (and the 
$CP$-conjugate process).
At the best-fit point in Eq.~\ref{BFPars}, this contribution is completely
negligible, despite being relatively enhanced over $\sigma_{bb}$ by a strange
quark parton distribution function (relative to a bottom parton distribution
function). It can contribute a few percent to the \ZP production
cross section at the edge of 
the 95$\%$~CL region where $\theta_{sb}=-0.15$.
We see therefore that the total \ZP production cross section
$\sigma$ -- still dominated by 
$\sigma_{bb}$ -- is quite insensitive to the
actual value of $\theta_{sb} \ll 1$ in 
our preferred-fit region in Eq.~\ref{rangePars}.
The cross section is, however, strongly
dependent on both $g_{Z^\prime}$ and $M_{Z^\prime}$, the latter via modifying
the energy fraction at which the parton
distribution functions are sampled.
There is also a non-negligible contribution to the cross section from
associated production with a 
$b$-quark or $\bar b$ anti-quark, as depicted in Fig.~\ref{fig:ffbar}. Associated
production is suppressed by the large \ZP mass reducing the phase
space, but is only singly suppressed by the bottom or anti-bottom parton
distribution function. We explicitly include this mechanism in the
computations outlined below.  

\begin{table}
  \begin{center}
  \begin{tabular}{|c|ccccc|}\hline
    mode & $b \bar b$ & $t \bar t$ & $e^+ e^-$ & $\mu^+ \mu^-$ & $\nu \bar
    \nu^{(\prime)}$ \\ \hline
       BR  & $\frac{2}{9}$ & $\frac{2}{9}$ & $\frac{2}{27}$ & $\frac{8}{27}$ &
    $\frac{5}{27}$ \\ \hline
  \end{tabular}
  \end{center}
  \caption{Branching ratios (BRs) into various fermion/anti-fermion pairs in the
    massless unmixed fermion approximation. \label{tab:decays}}
  \end{table}
In the massless fermion approximation, the \ZP decays with a total
width
\begin{equation}
  \Gamma = \frac{9}{8 \pi} g_{Z^\prime}^2 M_{Z^\prime}, 
  \end{equation}
  with branching ratios shown in Table~\ref{tab:decays}~\cite{Allanach:2024jls}.
  To remain in the perturbative regime, the \ZP width should not be too
  broad. Requiring $\Gamma / M_{Z^\prime} < 1/2$, we obtain $g_{Z^\prime} < 2
  \sqrt{\pi}/3 = 1.18$. Combining this with the lower limit on
  $g_{Z^\prime}/M_{Z^\prime}$ from Eq.~\ref{rangePars}, we see that
  $M_{Z^\prime} < 39$~TeV from the combination of perturbativity and the global fit of the
  \model\ Model's parameters. The lower the value of $M_{Z^\prime}$, the more
  perturbative the region of parameter space preferred by the global fit. 
  In what follows, we shall demarcate the $\Gamma_{Z^\prime}/M_{Z^\prime} \leq
  0.1$ region, where we may expect our predictions (which are based on
  perturbation theory) to be very accurate. Further away from this region,
  essentially at higher values of $g_{Z^\prime}$, the accuracy of our
  predictions and constraints will progressively degrade.

\subsection{Model scope \label{sec:scope}}
  
At 
  one-loop order, the renormalisation group equation of the $U(1)_X$ gauge coupling is 
  \begin{equation}
    \frac{d g_{Z^\prime}}{d \ln \mu} = \frac{g_{Z^\prime}^3}{16 \pi^2}
    \left[\frac{2}{3}\sum_{\psi} X_\psi^2 + \frac{1}{3}
      \sum_\phi X_\phi^2\right] = \frac{65}{48 \pi^2} g_{Z^\prime}^3, \label{eq:rge}
    \end{equation}
  where the sums are over all Weyl fermions $\psi$ of $U(1)_X$-charge
  $X_{\psi}$, over all complex scalar fields $\phi$ of $U(1)_X$-charge
  $X_{\phi}$ and where $\mu$ is the $\overline{MS}$ renormalisation scale. In order to attain the second equality in  Eq.~\ref{eq:rge} we
  have substituted 
  the charges of the fermions from 
  Table~\ref{tab:charges}, along with that of the flavon. Solving this equation, 
  we obtain a Landau pole in $g_{Z^\prime}$ at a scale
  \begin{equation}
    \Lambda_{LP} = M_{Z^\prime} \exp \left[ \frac{48 \pi^2}{130
        g(M_{Z^\prime})^2}\right].
    \label{eq:LP}
    \end{equation}
  Substituting the perturbativity bound on $g_{Z^\prime}$ from above, we
  obtain $\Lambda_{LP} \geq 14 M_{Z^\prime}$. One should require a new effective field theory
  to take over at a scale $\Lambda < \Lambda_{LP}$ in order to retain
  perturbative control. Further from the non-perturbative limit,
  $\Lambda_{LP}$ will increase significantly and there can be more separation
  between the \model\ Model and its ultra-violet completion. 

Generically, the gauge bosons of $U(1)_X$ kinetically mix with the hypercharge
gauge bosons.
The $U(1)$ kinetic  terms in the Lagrangian density appear as
\begin{equation}
  {\mathcal L}_{kin}^{U(1)} = -\frac{1}{4} X_{\mu \nu}X^{\mu\nu} -\frac{1}{4} B_{\mu\nu}B^{\mu \nu}
    -\frac{1}{2}\chi B_{\mu\nu} X^{\mu\nu},
\end{equation}
where $B_{\mu\nu}$ is the field strength of the hypercharge gauge boson field
and $X^{\mu\nu}$ is the field strength of the $X$ gauge boson field.
$\chi \neq 0$ results in the \ZP acquiring an additional component to its
coupling~\cite{HOLDOM1986196} $\propto g^\prime Y \chi$ to 
all fields, where $Y$ is the hypercharge of the field in question and
$g^\prime$ is the hypercharge gauge coupling.

In the present paper, we shall assume that $\chi$ is negligible. This could result
from the model being derived from some more unified higher-scale effective field theory which
has a semi-simple gauge Lie algebra. At this higher scale $\Lambda$,
kinetic mixing vanishes. Renormalisation group running then generically re-introduces non-zero kinetic mixing through loop corrections, however this would be suppressed by a loop factor. Indeed to leading-log order one obtains,
\begin{equation}
  \chi(M_{Z^\prime}) = \frac{g_Y g_{Z^\prime}}{16 \pi^2} \sum_\psi (Y_\psi
  X_\psi) \log \left( \frac{\Lambda}{M_{Z^\prime}} \right).  \label{chi}
\end{equation}
Substituting in $Y_\psi$ and $X_\psi$ from Table~\ref{tab:charges} and
$g_Y=0.35$ evaluated at $M_Z$~\cite{ParticleDataGroup:2022pth} in the SM into
Eq.~\ref{chi}, we obtain
$\chi = 0.018\ g_{Z^\prime} \log (\Lambda / M_{Z^\prime})$. In a
similar model (the malaphoric \ZP model~\cite{Allanach:2024ozu}), $\chi>0.01$ has a
non-negligible effect upon the LHC search limits, see Fig.~3 of
Ref.~\cite{Allanach:2024nsa}.  
Substituting this in, we find that for $g_{Z^\prime} \log (\Lambda /
M_\ZP) > 0.6$, one should 
recalculate the various bounds and sensitivities in the present paper
including the effects of kinetic mixing. We leave such a calculation to a
future publication. We note that the present analysis will be valid provided
$\Lambda$ is not too far above $M_{Z^\prime}$, depending upon the value of
$g_{Z^\prime}$ (unless there is a cancellation in $\chi$ between the
tree-level and loop contributions). We remark that this condition is compatible
with both $\Lambda < \Lambda_{LP}$ and Eq.~\ref{eq:LP}.

  Another limitation of the analysis in the present paper is that we have
  \emph{not} included the 
  effects of a propagating flavon. Thus, the level of effective field theory
  that we are considering is where the flavon has been integrated out and
  has a mass much greater than $M_{Z^\prime}$. We also neglect
  flavon-Higgs mixing: this occurs at a technically natural
  point\footnote{Here, by `technically natural', we mean that if the parameter
  is set to zero at some scale, renormalisation group evolution preserves its
  zero value at other scales.} where the
  potential coupling in the scalar potential $\lambda_{|\theta|^2 |H|^2}
  |\theta|^2 |H|^2$ has a zero coupling, i.e. $\lambda_{|\theta|^2 |H|^2}=0$.

  \section{Re-casting LHC Searches \label{sec:recast}}
A number of direct searches for new spin-1 resonances have been performed by
both the ATLAS and CMS collaborations at the LHC in various different assumed
final states. To date, no statistically significant signal has been observed
in any channel. Instead, upper limits have been placed on the  \ZP
production cross section multiplied by its branching fraction to the final
state in question as a function of the invariant mass of the resonance,
$M_{Z^\prime}$.  Of most relevance to the \model\  Model are searches
targeting di-lepton final states: $\mu^+ \mu^-$, $e^+ e^-$  for which the
branching fraction is expected to be appreciable.
Table~\ref{tab:decays} confirms that there is also a sizeable partial width for
\ZP decays into $b \bar b$ and $t \bar t$. 
We initially recast both an ATLAS \cite{ATLAS:2019fgd} and CMS~\cite{CMS:2019gwf} search for di-jet resonances, in addition to an ATLAS search for $t \bar t$ resonances in fully hadronic states \cite{ATLAS:2020lks}, following the re-interpretation procedures outlined in Ref.~\cite{Allanach:2021bbd}, to cover these channels. 
However, we found that due to the additional  
experimental challenges presented by such final states, these searches were not
sensitive to the parameter space of interest. 
We thus omit them from the rest of our discussion.
We have not re-cast LHC monojet searches such as Ref.~\cite{CMS:2021far}, which would be relevant for (for example)
 $Z^\prime (\rightarrow \nu \bar \nu)$ production plus initial state radiation. Such searches are expected to
be far less sensitive than di-muon or di-electron bump-hunts, due to both higher backgrounds and a lower signal cross section, which is suppressed by requiring additional hard radiation of an initial-state jet.

In this section we reinterpret various direct LHC searches using Run II data to place bounds on the \model\ \ZP boson over the two-dimensional parameter space spanned by $M_{Z^{\prime}}$ and $g_{Z^{\prime}}$. 
The generic procedure that we follow for each experimental search is as
follows. To start with we compute an estimate of the
relevant observable within the \model\ Model for each point in the
($M_{Z^{\prime}},g_{Z^{\prime}})$ parameter space, fixing $\theta_{sb}$
according to Eq.~\ref{BFPars}.
As Eq.~\ref{approx} and the surrounding discussion mentions, the \ZP
production cross section is insensitive to the precise value of this parameter. 
We  then compute the ratio $\mu$ of the 95\%
CL experimental upper bound on the experimental observable to our theoretical
prediction of the same quantity at each point in parameter space. A grid interpolation over the ($M_{Z^\prime}, g_{Z^{\prime}}$)  parameter plane is then used to determine the contour at $\mu = 1$. This corresponds to the upper limit on the model at the 95\% CL, which we then display on our plots.  

We also provide a rough assessment of the reach of the HL-LHC, assuming an integrated luminosity of $L =
3000$ fb$^{-1}$. To do this we use that for fixed centre-of-mass energy,
the number of both signal and background events are predicted to scale with
the integrated luminosity of the run.
Since the number of expected background events has an uncertainty $\propto 
\sqrt{L}$, the signal sensitivity  
is therefore expected to scale (in terms of the number of sigma, in the Gaussian
regime) as $L/\sqrt{L} = \sqrt{L}$. To provide a
forecast of the sensitivity, we compute the ratio $\tilde{\mu}$ of the
\textit{expected} 95\% CL experimental upper bound on
the observable in question to our  theoretically calculated result, and, once
again using a grid interpolation, solve for the contour at $\tilde{\mu} =
\sqrt{3000/L_0}$ where $L_0$ denotes the integrated luminosity in fb$^{-1}$ of
the search in question. Note that our theoretical predictions for the observables are computed assuming a centre-of-mass energy of $\sqrt{s} = 13$~TeV, 
appropriate for the Run II searches that we recast. The HL-LHC will, however, operate at a higher centre-of-mass energy, although the precise value of this is yet to be finalised. Since an increase
in run energy is expected to improve the signal sensitivity, our predictions
should be viewed as conservative. Given that the details of experimental
analyses are likely to differ between runs in any case, we deem this approximation
sufficient for our purposes.  

For the searches in question the relevant experimental observable corresponds
to  either a fiducial cross section (accounting for experimental acceptance)
or the total $Z^{\prime}$  production cross section, each multiplied by the
relevant branching ratio.  To compute the observables we rely on a Universal
FeynRules Output ({\tt UFO})~\cite{Degrande:2011ua} file\footnote{This file
can be found in the ancillary information attached to the {\tt arXiv} preprint
version of this paper.} from
Ref.~\cite{Allanach:2024jls}, which was created by 
implementing the \model\ Lagrangian in {\tt FeynRules}. This is imported into
{\tt MadGraph5 aMC@NLO v2.9.24}~\cite{Alwall:2014hca} which is then used to
simulate tree-level $Z^{\prime}$  production followed by decay into each
channel: $pp \rightarrow Z' \rightarrow X\bar{X} $ with $X \in \left\{ e^-,
\mu^- \right\}$ for a centre-of-mass energy of $\sqrt{s} = 13$~TeV.   We
explicitly include the associated production of the $Z^{\prime}$ boson
alongside a $b$-jet at the matrix element level. We use a 5-flavour scheme for
all of our computations in which the $b$-quark is included in the proton and
jet definitions\footnote{To implement this consistently we set the mass and
Yukawa couplings of the $b$  quark to zero, thus matching those of the light
quarks, via a {\tt UFO} restriction file.} and use the {\tt NNPDF2.3LO}
parton distribution functions (PDFs)~\cite{Ball:2012cx}. Our computations
incorporate tree-level off-shell contributions, but neglect interference with
SM background processes.
This should be accurate for a narrow resonance, but could affect the region
where the \ZP width-to-mass ratio $\Gamma/M_{Z^{\prime}}$ is large. To this end we shall delineate the
region where $\Gamma/M_{Z^{\prime}}$  larger than 0.1 on our plots. 
The simulated events are subsequently passed to {\tt
  Pythia 8.313}~\cite{Bierlich:2022pfr} to simulate parton showering, initial
state radiation and hadronisation. We do not include multi-parton interactions
in these simulations, which were found to have negligible impact on the
experimental observables of interest. To avoid double counting events with
final state jets that originate from the matrix element computation with those
from the parton shower, we match up to one jet using the MLM
procedure~\cite{Mangano:2006rw,Alwall:2007fs} in MadGraph~\cite{Alwall:2014hca} with a matching scale  (implemented
via the {\tt xqcut} parameter) of 40~GeV. This value was selected in order to
ensure smooth differential jet distributions and cross sections that are
robust under small changes to {\tt xqcut}.

To approximate the effect of higher-order QCD corrections to the \ZP
production rate, we could have rescaled our matched leading-order
cross section by a constant 
next-to-leading-order (NLO) QCD $K$-factor, $K\simeq 1.3$. The QCD corrections to
the production of a colour-singlet neutral vector are controlled by the colour
charges of the initial-state (anti-)quarks and are, to a very good approximation,
independent of the boson's couplings. $K$ may therefore be taken from the
extensively studied case of high-mass neutral Drell-Yan
production~\cite{Fuks:2007gk}.
We
have not applied this factor, noting that the $\mathcal{O}(\alpha_s)$ real-emission process $gb\rightarrow
Z^\prime b$, which dominates the NLO correction, is already present in our
MLM-matched sample. Also, note that the search based on $b$-jet
multiplicity -- the CMS di-muon-with-additional-$b$-jets analysis of
Sec.~\ref{sec:cmsdimuonb} -- may in principle carry a slightly different effective
$K$-factor, since the additional QCD radiation feeds directly into the $b$-jet
counting.

The published limits of the searches we recast implicitly account for detector inefficiencies, removing
the need for a full detector simulation. We nonetheless pass the {\tt
  Pythia}~\cite{Sjostrand:2014zea}
output to {\tt Delphes-3.5.0}~\cite{deFavereau:2013fsa} to perform jet-clustering and simulate
$b$-tagging and muon-isolation where necessary. Jet clustering is implemented
via the {\tt fastjet-3.5.1}~\cite{Cacciari:2011ma} plugin using the anti-$k_t$ algorithm~\cite{Cacciari:2008gp} with a distance
parameter of $R=0.4$. The reconstructed jet and lepton objects are written to
a {\tt ROOT} file and ordered in terms of transverse momentum. We shall henceforth refer to the $i^\text{th}$ jet and lepton in the corresponding ordered collections as $j_i$ and $l_i$, respectively, with $i=1$ denoting the leading object.

The various searches considered differ in whether detector acceptance is folded into the published experimental bounds. To enable the same simulated events to be used for multiple analyses, we do not impose any phase-space cuts at the parton-level. In this way we may use the matched cross sections as our estimates for the total cross section times branching ratio $\sigma \times BR$, to be used for searches which correct for detector acceptance prior to reporting. For searches that do not correct for this, we multiply $\sigma \times BR$ by an estimated acceptance fraction $f$, which we obtain by applying the phase-space cuts used in the experimental analysis to the {\tt Delphes} output using a {\tt ROOT} python macro. 

Finally we note that some of the searches to be recast report constraints for several different values of the $Z^{\prime}$ width-to-mass ratio $\Gamma/M_{Z^{\prime}}$. We denote this set by $\{W_1,W_2, \ldots ,W_n\}$ ordered such that $W_j > W_i$ for $j > i$, with $W_1$ corresponding to the $\Gamma \rightarrow 0$ limit.  We denote the corresponding experimental bounds by $\{B(W_1,M_{Z^\prime}),B(W_2,M_{Z^\prime}), \ldots , B(W_n,M_{Z^{\prime}})\} $. For a parameter space point with $z \equiv \Gamma/M_{Z^{\prime}}$ satisfying $W_p < z < W_{p+1}$ we determine the experimental bound by linearly interpolating  between $\ln(B(W_p,M_{Z^\prime}))$ and  $\ln(B(W_{p+1},M_{Z^\prime}))$. Explicitly 
\begin{equation}
B(z,M_{Z^\prime}) =  B(W_{p},M_{Z^\prime})\left(\frac{B(W_{p+1},M_{Z^\prime})}{B(W_{p},M_{Z^\prime})}\right)^{\left(\frac{z - W_{p}}{W_{p+1} -W_{p}}\right)}. \label{eq:widthinterp}
\end{equation} For $z > W_n$, we set $B(z,M_{Z^\prime})$ = $B(W_n,M_{Z^\prime})$ rather than performing an extrapolation. It should be noted that the \model\ model enters the non-perturbative regime when $\Gamma/M_{Z^{\prime}} \sim \mathcal{O}(1)$. In this region our results, which are based on perturbative calculations, become increasingly  unreliable. We explicitly mark the contour $\Gamma/M_{Z^{\prime}} = 0.1$ to indicate a high level of confidence in the results for the parameter space below this (i.e. with smaller $g_\ZP$).

In the following subsections we provide further details regarding each of the searches used and the approximations made in our re-interpretation procedure. Our approach closely follows that deployed in Ref.~\cite{Allanach:2021bbd} for a different $Z^{\prime}$ model.  

\subsection{ATLAS di-lepton search}
We first re-interpret a search for high-mass di-lepton resonances
\cite{ATLAS:2019erb}, incorporating 139 fb$^{-1}$ of $pp$ LHC Run II collision
data collected at $\sqrt{s}=13$~TeV with the ATLAS detector. The search found
no significant excess and placed bounds on a fiducial cross section times
branching ratio as a function of the $Z^{\prime}$ mass for various different
values of the $Z^{\prime}$ width-to-mass ratio up to a maximum value of
0.1. To estimate the acceptance fraction for di-muon final states, we select
the number of events including a $\mu^+ \mu^{-}$ pair each with transverse
momentum magnitude $p_T > 30$~GeV, absolute pseudo-rapidity $|\eta| < 2.5$ and
invariant mass $m_{\mu \mu }> 225$~GeV.
Each muon is required to be isolated in that the  scalar sum of tracks with $p_T > $ 1~GeV within a  
cone of size $\Delta R = \sqrt{(\Delta \eta)^2 + (\Delta \phi)^2}$ to the
muon must be less than 6\% of the transverse momentum of the candidate muon
($p_T(\mu)$), where $\Delta \eta$ and $\Delta \phi$ are the difference in the 
pseudo-rapidity and azimuthal angle of the track and muon. The value of
$\Delta R$ for a given muon candidate is taken to be the minimum of 0.3 and
the ratio 10~GeV/$p_T$. We implement this by modifying the default isolation
module in {\tt Delphes}. 

For the di-electron search we estimate the acceptance fraction by enumerating
the number of events containing a pair of oppositely charged electrons with
invariant mass $m_{e e }> 225$~GeV. Each electron is additionally required to
have a transverse momentum $p_T > 30 $~GeV and absolute pseudo-rapidity
$|\eta| < 2.5$ excluding the region $1.37 < |\eta| < 1.57$ which corresponds
to the transition between the barrel and the end-cap. The electrons must also pass the `gradient' isolation working point defined in
Ref.~\cite{ATLAS:2019jvq} which is characterised by a $p_T$ dependent isolation efficiency: $\epsilon_{\rm iso} = 0.1143 p_T(e)/\text{GeV} + 92.14$\% (capped at a
maximum of 100$\%$). We implement this within the {\tt ROOT} macro: for each electron we compute $\epsilon_{\rm iso}$ then draw a random number $\xi$ from a uniform distribution between 0 and 100. The electron is deemed isolated if $\xi < \epsilon_{\rm iso}$.

Although we recast the electron and muon searches independently, the exclusion bound and projected HL-LHC sensitivity shown in Fig.~\ref{fig:LHC} correspond to a combined result obtained by selecting the most stringent constraint out of the two channels at each point in parameter space.

\subsection{ATLAS di-lepton search with one or no $b$-tagged jets}
 We next look at an ATLAS search for new resonances in di-lepton final states with either zero or one $b$-tagged jets in the full Run II dataset comprising 139 fb$^{-1}$ of $pp$ collision events at $\sqrt{s} = $ 13~TeV \cite{ATLAS:2021mla}. The analysis reports bounds on the cross section times branching ratio including the acceptance fraction and (where relevant) the $b$-tagging efficiency.  To account for $b$-tagging efficiency, we use the default $b$-tagging module in {\tt Delphes} to tag $b$ jets with a fixed efficiency of 77\%. 
 
 We obtain an estimate of the acceptance fraction by applying phase-space cuts to match the experimental selection criteria. Specifically we require any jet to have $p_T >$ 30~GeV and $|\eta|< 2.5$. Similarly, muon candidates need to have  $p_T >$ 30~GeV and $|\eta|< 2.5$. Electrons must have energies exceeding 30~GeV and $|\eta|< 2.47$ excluding the region $1.37 <|\eta|< 1.52$. 

 Overlapping jet and leptonic objects are sequentially removed according to
 the following  algorithm, which we implement in the {\tt ROOT} macro. First,
 jets within $\Delta R < 0.2$ to an electron are removed. Then electrons
 within $\Delta R <$ 0.4 from a remaining jet are removed. Finally, if a jet
 is within $\Delta R < 0.04 + 10 \textnormal{~GeV}/p_T(\mu)$ of a muon object,
 it is removed provided it has no more than 2 constituent tracks (charged
 leptons and charged hadrons) with transverse momentum $p_T > 0.5$~GeV, otherwise
 the muon is removed. 

Upon completion of this procedure, for both the muon and electron searches,
events which contain an opposite sign same flavour pair of leptons are
separated into those with no $b$-tagged jets ($b$-veto) and those with exactly
one $b$-tagged jet ($b$-tag). The transverse momentum of
the leading lepton $p_T(l_1)$ is required to exceed 65~GeV to ensure trigger
selection. 

The experimental bounds for this search are reported as a function of the
minimum invariant mass of the di-lepton system, $m_{ll}$.  For each point in the
($M_{Z^\prime},g_{Z^{\prime}}$) parameter space we apply the experimental
limit evaluated at $m_{ll} = M_{Z^\prime} - 2 \Gamma_{Z^{\prime}}$, to obtain
a conservative estimate for the exclusion bound.  Given that the $Z^{\prime}$
width is small over the majority of parameter space in question, the precise
mapping used here does not have a significant impact on the derived
constraints.    

Once again, our plots display the most constraining bound/forecast of these
four distinct searches ($e$-tag, $e$-veto, $\mu$-tag, $\mu$-veto) at each
point in parameter space. For both the Run II bounds and our forecasts for the
HL-LHC, we find that the muon $b$-veto search performs the best across the
entire parameter space under consideration.

\subsection{CMS di-lepton}
We recast a CMS search for new narrow resonances in di-lepton ($l^{+}l^{-}$ for $l$ $\in$ $\{e,\mu \}$)  final states using the full set of Run II $pp$ collision data at $\sqrt{s} = 13$~TeV. The analysis found no significant excess and placed  upper limits on~\cite{CMS:2021ctt}
\begin{equation}
\frac{\sigma \times BR(Z^{\prime} \rightarrow l^+ l^-)}{\sigma_Z \times BR(Z \rightarrow l^+l^-)}\times 1928 \textnormal{~pb}. 
\end{equation}
Here $\sigma_Z$ denotes the total production cross section of the $Z^0$ boson
and the SM prediction for the denominator is 1928~pb. We estimate the
numerator for each channel using our matched muon and electron computations
with no additional selection cuts applied. For the denominator we take
analogous matched computations for the $Z^0$ boson. The experimental
analysis reported bounds for $Z^{\prime}$ width-to-mass ratios of 0\%, 0.6 \%,
3 \%, 5\% and 10 \% which we interpolate between as described above. We
extract bounds (or HL-LHC projections) for the electron and muon channels separately, then display the most constraining at each point in parameter space in our plots. In both our Run-II bounds and HL-LHC forecasts, this corresponds to the di-muon channel across the entirety of the scanned parameter space. 

\subsection{CMS di-muon with additional $b$-jets}\label{sec:cmsdimuonb}
We finally consider a CMS search for high-mass di-muon resonances produced in association with $b$-jets \cite{CMS:2023byi}. The study uses 138 fb$^{-1}$ of $pp$ collision data at $\sqrt{s}$ = 13~TeV and, having been designed with $Z^{\prime}$ models that modify $b\rightarrow s l^+ l^-$ transitions in mind, is expected to be particularly sensitive to the \model\ Model $Z^{\prime}$. No significant excess was found and exclusion limits were placed on the number of events passing the analysis selection containing one or more $b$-tagged jet as a function of $M_{Z^\prime}$ for various values of $f_{2b}$ between 0 and 1, where $f_{2b}$ denotes the fraction of events passing selection that have at least two $b$-tagged jets. 

To pass the selection, events must contain a pair of oppositely charged, isolated muons originating from the primary vertex\footnote{An object is considered consistent with originating from the primary vertex if its transverse impact parameter is less than 2 mm and its longitudinal displacement along the beam-line is less than 1 mm.} with $p_T > $ 53~GeV and $|\eta| < 2.4$. A muon is considered isolated if the scalar $p_T$ sum of all charged tracks, excluding that of the muon, within a cone of $\Delta R = 0.3$ relative to the muon track and within 2 mm of the primary vertex along the beam direction is less than 5~GeV and less than 5 \% of the $p_T$ of the muon.  Events containing additional muons with  $p_T > $ 10~GeV and $|\eta| < 2.4$ satisfying the same isolation criteria are vetoed.  
Events with isolated electrons with $p_T > 10$~GeV and $|\eta|<2.5$ are
vetoed. 

To remove the main experimental backgrounds, primarily tau decays, a second
veto is applied to exclude events containing additional isolated charged tracks
(muons, electrons and charged hadrons) which originate from the primary vertex
and have  $|\eta| < 2.5$ (2.4 in the case of muons), and $p_T \geq 5 $ (10)
GeV for leptons (charged hadrons). Such tracks are deemed isolated if the
scalar $p_T$ sum of other charged tracks within a cone of $\Delta R = 0.3$
with respect to the track in question is less than 20\%  of the isolated track
$p_T$ for leptons and $10\%$ for charged hadrons. We determine whether tracks
are isolated by modifying the default isolation module in {\tt Delphes} to
realise these criteria.  

Jets are required to have $p_T > 20$~GeV,
$|\eta|<2.5$ and be separated by $\Delta R > 0.4$ from the leading muon and anti-muon
candidates. At least one $b$-jet is required to satisfy tight identification
criteria corresponding to a 58\% tagging efficiency, and a 0.1 \%
misidentification probability. All other $b$-quark jets are identified using
relaxed requirements corresponding to a 67 \% tagging efficiency and $\sim$1
\% misidentification probability. To implement this, we tag every jet under
loose requirements using the default b-tagging module in {\tt Delphes}. The
probability that a jet which satisfies the loose identification criteria also
satisfies the high identification criteria is $P_{h|l} = 58/67$. Hence, in an
event with $n_b$ $b$-jets satisfying the loose criteria, the probability that
none of them satisfy the tight identification procedure is $P_E = (1-
P_{h|l})^{n_b}$. For each event, we generate a random number, $\zeta_b$ from a
uniform distribution between 0 and 1, and reject the event if $\zeta_b <P_E$.

The minimum invariant mass of any muon-$b$-jet pairing, min($m_{\mu b}$) (where the muons correspond to those forming the $\mu^{+}\mu^{-}$ pair) is required to exceed 175~GeV. This threshold is above the $t$ quark mass and is imposed to remove backgrounds from top decay.  

During 2018, a sector of the hadronic calorimeter covering $-3.2 < \eta <
-1.3$ and $-1.57 < \phi <-0.87$ was not operational for data taking, resulting
in the mis-reconstruction of hadronic jets in the $\eta, \phi$ plane. Events
collected during this period, corresponding to an integrated luminosity of 39
fb$^{-1}$,  were rejected if jets or electrons were found in this region, with
the upper and lower boundaries for jets extended by 0.2 in both $\phi$ and
$\eta$ to account for the size of jet cones. We apply these cuts to a fraction
$F = 39/138$ of simulated events. The events to which to apply these criteria
are
selected randomly by drawing a number $\zeta_F$ from a uniform distribution on
(0,1) and implementing the cuts if $\zeta_F < F$. We do not attempt to apply
any of the other cuts undertaken in the experimental analysis since they were
aimed at removing backgrounds which are not
produced in our signal simulation and are
thus negligible.

The number of events passing all of the above selection cuts is multiplied by
$0.9$ to account for the (approximate) experimental trigger efficiency, yielding
our estimate of the expected signal yield in the di-muon-plus-$b$ signal region:
\begin{equation}
  N_s(M_{Z^\prime},g_{Z^\prime}) = 0.9\,\times\, \sigma\times
  BR(Z^\prime\rightarrow\mu^+\mu^- + j_b)\,\times\, {\mathcal L}\,\times\,{\mathcal A}~.
  \label{eq:Ns}
\end{equation}
Within this expression, ${\mathcal L}=138$~fb$^{-1}$ is the integrated luminosity of the
search~\cite{CMS:2023byi}, and ${\mathcal A}$ is the fraction of generated events that survive the full
selection chain described above, including the stochastic tight-tag acceptance
encoded in the rejection on $\zeta_b<P_E$ and the HCAL-hole veto encoded in the
rejection on $\zeta_F<F$. We explicitly write $j_b$ in the expression for the branching fraction to indicate that the final state must include at least one $b$-tagged jet.

The experimental limits are reported as a function of $M_{Z^\prime}$ for several
representative values of the double-$b$-tag fraction $f_{2b}$ between 0 and 1, defined as the
fraction of selected signal events that contain at least two $b$-tagged jets. To
confront the \model\ Model with these limits, we must therefore predict $f_{2b}$
at each point in parameter space. To this end we count, in every retained event, 
the number $n_b$ of jets that satisfy the loose ($67\%$-efficiency) $b$-tagging
requirement, as modelled by {\tt Delphes}; this multiplicity automatically
includes the small contamination from mis-tagged charm- and light-flavour jets.
Recalling that an event is retained only if at least one such jet is promoted to
a tight tag (the rejection on $\zeta_b<P_E$ introduced above), we classify a
retained event as \emph{double-tagged} if $n_b\geq 2$ and as \emph{single-tagged}
if $n_b=1$. Denoting by $N(n_b\geq 2)$ and $N(n_b\geq 1)$ the number of retained
events with at least two and at least one loose $b$-tagged jet respectively, the
predicted double-$b$-tag fraction is
\begin{equation}
  f_{2b}^{\rm th}(M_{Z^\prime},g_{Z^\prime}) =
  \frac{N(n_b\geq 2)}{N(n_b\geq 1)}.
  \label{eq:f2b}
\end{equation}
We find $f_{2b}^{\rm th}$ to be essentially independent of $g_{Z^\prime}$, since
an overall rescaling of the coupling multiplies all rates equally. Further, it varies
only mildly with $M_{Z^\prime}$ across the scanned range. This is because it is governed by the
$b$-jet content of the signal, which is dominated by $b\bar b\rightarrow
Z^\prime$ production, supplemented by associated $b\,Z^\prime$ production
(cf.\ Fig.~\ref{fig:ffbar}).  The relative weighting of these two processes shifts slowly with $M_\ZP$ as the
sampled parton momentum fractions change.

Denoting the discrete set of double-$b$-tag fractions for which experimental
bounds are supplied by $\{f_1,f_2,\ldots,f_m\}$, ordered such that $f_j>f_i$ for
$j>i$, and the corresponding upper bounds on the signal yield by
$\{B(f_1,M_{Z^\prime}),B(f_2,M_{Z^\prime}),\ldots,B(f_m,M_{Z^\prime})\}$, we
determine the bound at the predicted fraction by
linear interpolation.
For $f_{2b}^{\rm th}$ satisfying
$f_p\leq f_{2b}^{\rm th}<f_{p+1}$, 
\begin{equation}
  B(f_{2b}^{\rm th},M_{Z^\prime}) = B(f_p,M_{Z^\prime}) +
  \frac{f_{2b}^{\rm th}-f_p}{f_{p+1}-f_p} [B(f_{p+1},M_{Z^\prime}) -
    B(f_p,M_{Z^\prime})    ].
  \label{eq:f2binterp}
\end{equation}
The $95\%$~CL exclusion then follows, as for the other searches,
from the ratio $\mu = B(f_{2b}^{\rm th},M_{Z^\prime})/N_s$ evaluated over the
$(M_{Z^\prime},g_{Z^\prime})$ plane, with the contour at $\mu=1$ delimiting the excluded
region.

\subsection{Results}
\begin{figure}
    \begin{center}
      \includegraphics[width=0.98 \textwidth]{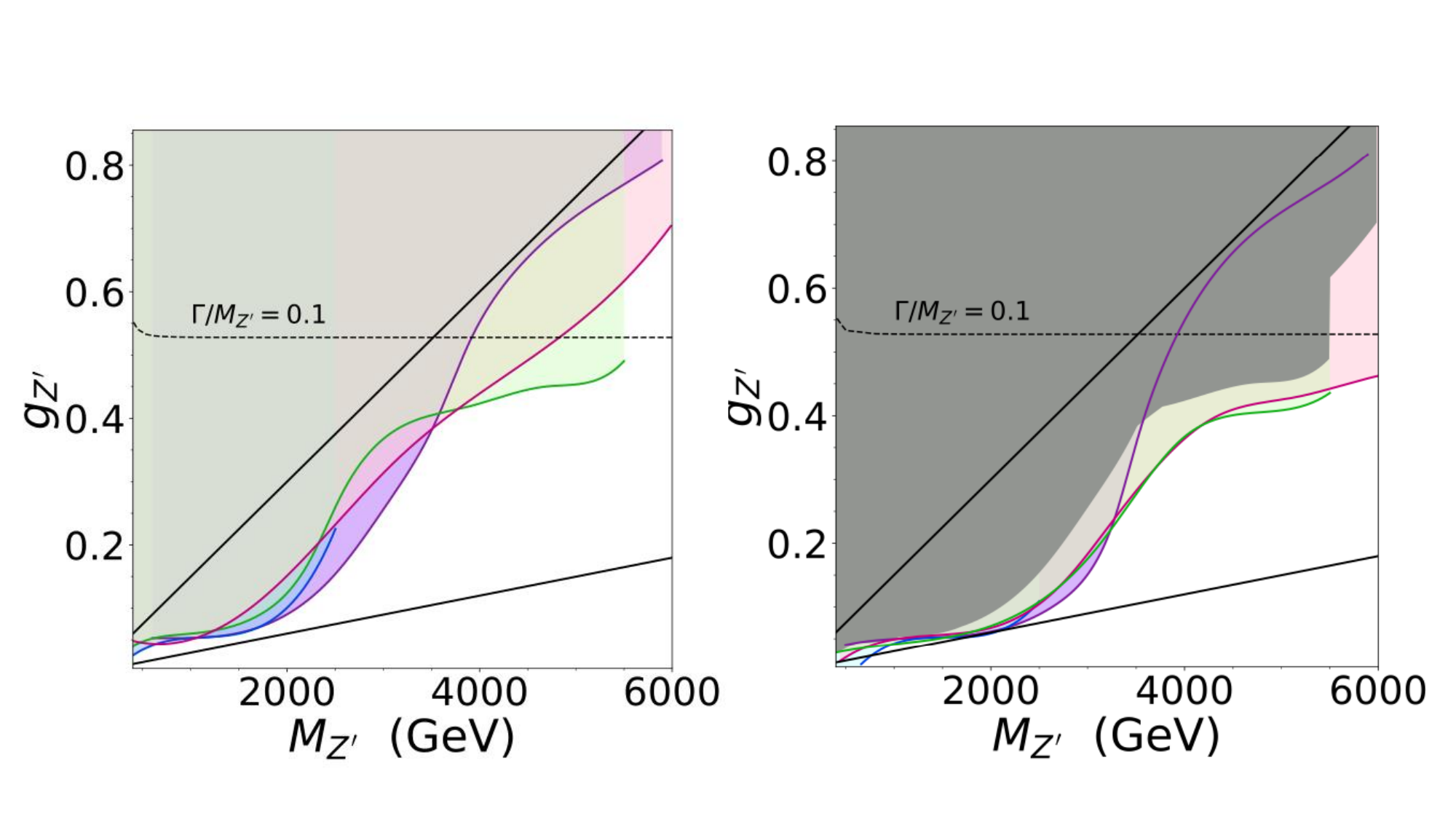}
      
    \begin{tabular}{ll}
    \swatch{xkcdLilac} ATLAS di-lepton $b$-tag/veto \cite{ATLAS:2021mla} & 
    \swatch{xkcdPalePink} ATLAS di-lepton \cite{ATLAS:2019erb} \\
    \swatch{xkcdPaleBlue} CMS di-muon with $b$ jets \cite{CMS:2023byi} &
    \swatch{xkcdPaleGreen} CMS di-lepton \cite{CMS:2021ctt}\\
    \end{tabular}
\end{center}
    \caption{\label{fig:LHC} (left panel) \ZP search constraints on the
    \model\ Model parameter space from Run II LHC searches. (right panel)
    expected sensitivity at the HL-LHC assuming 3000~fb$^{-1}$ of integrated
    luminosity. In the left (right) panels, the coloured regions delineate the
    95$\%$~CL exclusion
    (expected sensitivity of the HL-LHC) for each of the re-interpreted searches. Above the black dotted line $\Gamma_{Z^\prime}/M_{Z^\prime}>0.1$. The parameter space between the straight radial black lines corresponds to the 95$\%$~CL preferred-fit
    IMPs from Ref.~\cite{Allanach:2023uxz}. 
    In the right-hand panel, the dark grey shaded region denotes  the union of the 95$\%$~CL exclusion regions from the Run II LHC searches in the left-hand panel. 
  }
\end{figure}

We display the current bounds on the parameter space of the \model\ Model from the various searches mentioned above in the left-hand panel of Fig.~\ref{fig:LHC}.
As described in the preceding section, with the exception of the CMS di-muon search with additional $b$-jets, we derived independent bounds from the electron and muon channels for each of the analyses considered above. For the ATLAS search for di-leptons with one or no $b$-tagged jets, we further obtained separate bounds from the $b$-tag and $b$-veto signal regions, in both the muon and electron channels. The coloured regions in the plot correspond to the union of the various 95 \% CL exclusions from each search. Other than the ATLAS di-lepton search where, due to statistical fluctuations in the observed bound, the recast di-electron bound dips below that from the di-muon in a number of narrow mass ranges, we find the di-muon channel to be more constraining across the entire range of $M_{Z^{\prime}}$ considered. This is to be expected since, as
Table~\ref{tab:decays} shows, we expect four times the signal rate into
di-muons.  For the ATLAS search with one or no $b$-tagged jets, we find the $b$-veto signal region to provide a stronger constraint than the $b$-tag signal region throughout parameter space. We attribute this to
the top background; as Fig.~3 of Ref.~\cite{ATLAS:2021mla} shows,
requiring a $b$-jet results in a larger top background in the signal region at
high di-lepton invariant masses, degrading the performance.

The parameter space preferred at the 95$\%$ CL by the fit to the $b\rightarrow s$ transition
observables corresponds to the region between the two black radial lines in
Fig.~\ref{fig:LHC}.  As shown in the left-hand panel, a significant fraction of this region is already excluded by existing LHC \ZP searches. An additional portion of
this parameter space is expected to be probed by upcoming HL-LHC searches, as
illustrated by comparing the coloured regions in the right-hand panel, which denote the estimated HL-LHC sensitivity, with the union of the existing LHC exclusion regions, shaded in dark grey.

We also see from the figure that there is effectively no 95$\%$~CL bound on
$M_{Z^\prime}$  from the LHC searches within the 95$\%$~CL preferred-fit
IMPs but that HL-LHC is expected to have sensitivity up to around
$M_{Z^\prime} \geq 2500$~GeV.
The figure also demonstrates that analyses with additional $b$-tags and
$b$-vetoes vary the sensitivity of each search somewhat, providing the leading bounds for $M_{\ZP} <$  3500 GeV, but tailing off in sensitivity more quickly the standard di-lepton searches moving to higher masses. 
We emphasise again here that the HL-LHC estimates shown in the right-hand plot are likely too conservative, given
that they are strictly for a centre-of-mass energy of 13~TeV. The
HL-LHC is likely to be run at a higher centre-of-mass energy (closer to
14~TeV) that should afford somewhat better sensitivity.

\section{LHC Differential Cross-Section Measurements\label{sec:obs}}

Many of the differential cross sections which have already been measured by the LHC experiments offer
potential sensitivity to the \ZP of the \model\ Model.
The \contur~\cite{Butterworth:2016sqg,Buckley:2021neu} application is designed to
allow a rapid check of detailed models against those measurements which are available in Rivet~\cite{Buckley:2010ar}.
We have simulated events using \herwig~\cite{Bellm:2015jjp}, based upon the same UFO files as the previous section, again without
applying any $K$-factors. 
The {\tt NNPDF2.3LO} PDFs are again used (the current \herwig default).
These events are then passed through \rivet~4~\cite{Bierlich:2024vqo} and analysed with \contur~3.1.4~\cite{CONTUR:2025yis}.
This involves injecting the potential \ZP contribution on top of the SM predictions for a wide range
of LHC measurements, and evaluating the ratio between the likelihood evaluated for the SM alone given the data,
and the likelihood for the SM plus beyond-the-SM contributions.
The $CL_S$ method~\cite{Junk:1999kv,Read:2002hq} is used to determine an exclusion limit;
the expected exclusion limit is also determined by setting the data to the SM values but preserving its uncertainties.
Additionally, as in the previous section, a crude estimate of the expected eventual sensitivity of the HL-LHC is evaluated, by scaling the
data uncertainties for all 13~TeV measurements by the square root of the ratio of the current integrated luminosity to the
anticipated 3000~fb$^{-1}$, emphasising again that
this is likely to be very conservative~\cite{Belvedere:2024wzg}.
The cross-section measurements considered in the analysis include di-jets~\cite{ATLAS:2024png}, top~\cite{ATLAS:2017cez,CMS:2021vhb}, $t\bar{t}b\bar{b}$~\cite{ATLAS:2018fwl,CMS:2023xjh} and missing transverse momentum (monojet)~\cite{ATLAS:2024vqf} measurements, which could in principle show 
sensitivity to decays of the \ZP to particles other than charged leptons.

\begin{figure}[t]
  \centering
  \includegraphics[width=0.95\linewidth]{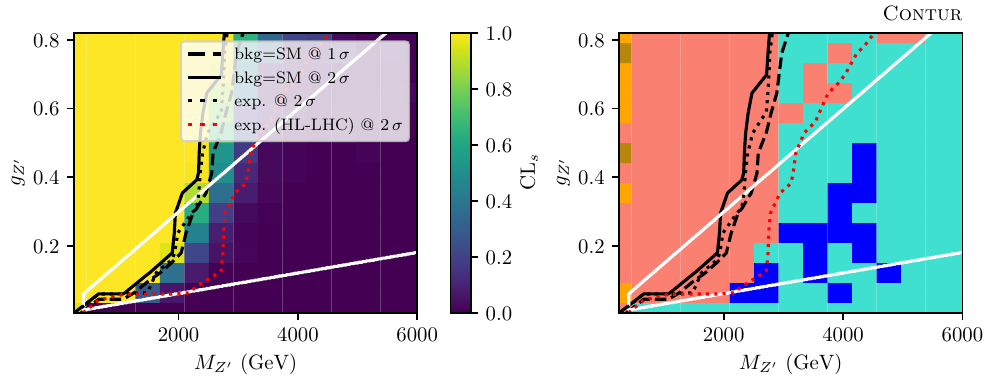}
  \begin{tabular}{lll}
    \swatch{orange}~$l^+l^-$+jet \cite{CMS:2022ubq} & 
    \swatch{turquoise}~$l_1l_2$+\MET{}+jet \cite{ATLAS:2024aht,ATLAS:2023gsl} & 
    \swatch{salmon}~high-mass Drell-Yan $ll$ \cite{ATLAS:2016gic,ATLAS:2019erb,CMS:2018mdl} \\
    \swatch{blue}~$l$+\MET{}+jet \cite{CMS:2021vhb,CMS:2016oae} &
    \swatch{darkgoldenrod}~$l^+l^-\gamma$ \cite{ATLAS:2022wnf,ATLAS:2019gey,ATLAS:2022wmu} 
  \end{tabular}
  \caption{\contur exclusion plot of $g_\ZP$ against $M_\ZP$. The left panel colour map indicates the
  combined exclusion of the measurements considered, while the colouring in the right panel indicates the most sensitive final state. In both panels, the solid black line indicates
    the 95\% exclusion and the dashed black line the 68\% exclusion.
    The dotted black line shows the expected 95\% exclusion, were measurements
    to exactly coincide with the SM predictions, and the dotted red line is an estimate of the HL-LHC sensitivity
    after 3000~fb$^{-1}$ of integrated luminosity.
    Between the straight radial lines is the 95$\%$~CL preferred region of
    fit from Ref.~\cite{Allanach:2023uxz} to $B$-meson decay and mixing data.
  For values of $M_{Z^\prime}$ that are too small, the fit of Ref.~\protect\cite{Allanach:2023uxz} is
inaccurate because it 
was described by SMEFT which is not accurate for calculating electroweak or
LEP2 results. We therefore only display the preferred-fit IMPs region for
$M_{Z^\prime}\geq 400$~GeV. }
  \label{fig:contur}
\end{figure}

The results are shown in Fig.~\ref{fig:contur}. While top production channels have some small sensitivity at high \MZP,
the only region excluded is below around 2.8~TeV, and this is excluded by the Drell-Yan measurement from CMS and the ATLAS
search already discussed in the previous section. Although
a very wide range of other final state measurements is also considered, there is still ample room for the \model\ Model to be viable
for \MZP above a~TeV or two, depending on the value of \GZP\@.
We note that the constraints from the Drell-Yan search are weaker than those shown in the previous section -- this is
due to the statistical treatment in \contur, which is intended for binned regions with significant cross sections and for which
the assumptions break down in the high-mass tail of the search analysis.

\section{Summary and Discussion \label{sec:conc}}

The \model\ Model explains some gross features of the fermion mass spectrum
and possesses a TeV-scale \ZP boson which can be both searched for directly and
affect LHC collider observables. 
We see from Figs.~\ref{fig:LHC} and~\ref{fig:contur} that
currently, direct \ZP
searches constrain the \model\ Model
more strongly than the various other differential cross-section
measurements included in \contur. Fig.~\ref{fig:LHC} shows that a
naive projection of the direct HL-LHC \ZP 
search sensitivity covers some parameter space of interest (i.e.\ where the
fit to $b \rightarrow s$ transition observables is ameliorated) that is
currently not excluded at the 95$\%$ CL, although the expected gains are a small fraction of the area of the remaining parameter space. 

It is clear from Ref.~\cite{Allanach:2023uxz} that models in which the \ZP couples to
di-electron pairs as well as di-muon pairs (in addition to coupling to
$b\overline{s}+h.c.$) are preferred in the fit to flavour (and other)
measurements. $3B_3-L$, i.e.\ thrice third-family baryon number minus lepton
number, is one notable possibility which (while outside of the $68\%$ CL and
because of which was discarded by Ref.~\cite{Greljo:2022jac}) is 
within the 95$\%$ CL and looks to be worthy of further attention. 
Other possibilities of well-fit models are discussed within 
Refs.~\cite{Greljo:2022jac,Allanach:2023uxz},
for example a model in which the \ZP may
seem initially to only couple to second
family leptons (the $B_3-L_2$ model) but which in fact couples to the other
families through order unity kinetic mixing~\cite{HOLDOM1986196} with the hypercharge gauge
boson\footnote{The phenomenology of such a model is arbitrarily close to a
model without kinetic mixing 
but in which a certain rational multiple of hypercharge is added onto the new $U(1)_X$
charges of the chiral fermionic fields~\cite{Allanach:2023uxz}.}. As discussed
in \S\ref{sec:scope}, we have neglected kinetic mixing in the present work.

Fig.~\ref{fig:LHC} displays the 95 \% CL exclusion bounds from various direct LHC searches for the \model\ Model, as well as a conservative
estimate of the further
sensitivity reach afforded by analysing data from the HL-LHC in the various
different channels. The concurrence of the various different channels in some
parts of parameter space will provide further tests of the Plan B Model. As shown in Fig.~\ref{fig:contur},
\contur provides bounds which agree with those from the direct searches towards the smaller
$M_{Z^\prime}$ values tested. 

In future it will be of interest to calculate the phenomenology of the
\model\ Model
including the effects of the flavon. Such an analysis is  appropriate for the \model\ Model
when the flavon mass is of the order of, or less than,
$M_{Z^\prime}$. In this instance, we       
expect \contur bounds from differential cross-section measurements to be stronger
than those calculated in the present paper
because experience has demonstrated that more complete models that consequently have more
complex phenomenology tend to be more constrained than very simple
ones (see, for example, Refs.~\cite{Amrith:2018yfb,Butterworth:2025szb}). Including the flavon in the phenomenology 
will notably lead to Higgs field-flavon field mixing~\cite{Allanach:2022blr}, affecting
the predictions of various Higgs (and other) differential cross sections and measurements. 

Differential cross-section measurements, unfolded for detector effects,
require a more painstaking experimental 
analysis than bump-hunts. In the LHC runs up until Run III, a large volume of
search parameter space has been covered by such bump hunt searches.
As the HL-LHC moves onward, however, not advancing the energy of collisions
significantly but 
instead increasing the already large integrated luminosity, we will start to reach
the law of diminishing returns for such searches.
One can already see evidence of this in
Fig.~\ref{fig:LHC}, where the increased luminosity of the HL-LHC affords
only a modestly increased sensitivity.
In our opinion, this helps motivate a gradual rebalancing of experimental analysis
methods as HL-LHC integrated luminosity increases, moving \emph{away} from
bump-hunts and \emph{towards} higher precision differential cross-section
measurements in channels motivated by direct searches, allowing 
us to find hints of new physics within them. In the case that no significant
deviations from the SM are found, 
differential LHC cross-section measurements bound the parameter space of potential
models beyond the SM and would form a
monolithic legacy, much as LEP has. 

\section*{Acknowledgements}
This work was partially supported by STFC HEP Consolidated grants
ST/T000694/1 and ST/X000664/1.  HB is support by the James Arthur Postdoctoral Fellowship at New York University. 
We are grateful to C.~Campagnari and M.~Masciovecchio for clarification of experimental details pertaining to the CMS di-muon search with additional $b$-jets~\cite{CMS:2023byi} and thank the Cambridge Pheno Working Group for helpful discussions.

\bibliographystyle{JHEP-2}
\bibliography{zp.bib}

\end{document}